# Spin gap evolution upon Ca doping in the spin ladder series $Sr_{14-x}Ca_xCu_{24}O_{41}$ by inelastic neutron scattering

G. Deng[a*], N. Tsyrulin[b], P. Bourges[c], D. Lamago[c], H. Ronnow[b], M. Kenzelmann[d], S. Danilkin[a], E. Pomjakushina[d], K. Conder[d]

[a] *Bragg Institute, ANSTO, New Illawarra Road, Lucas Height, NSW 2233, Australia*

[b] *Laboratory for Quantum Magnetism, Ecole Polytechnique Federale de Lausanne, CH-1015 Lausanne, Switzerland*

[c] *Laboratoire Léon Brillouin, CEA-CNRS, CE-Saclay, 91191 Gif sur Yvette, France*

[d] *Laboratory for Development and Methods, Paul Scherrer Institute, CH-5232 Villigen, Switzerland*

**Abstract**

The spin gap evolution upon Ca doping in $Sr_{14-x}Ca_xCu_{24}O_{41}$ was systematically investigated using inelastic neutron scattering. We discover that the singlet-triplet spin gap excitation survives in this series with *x* up to 13, indicating the singlet dimer ground state in these compounds. This observation corrects the previous speculation that the spin gap collapses at *x*~13 by the NMR technique. The strong intensity modulation along $Q_H$ in *x*=0 gradually evolves into a Q-independent feature in *x*>11. This could be attributed to the localized Cu moment magnetism developing into an itinerant magnetism with increasing *x*. It is a surprise that the spin gap persists in the normal state of this spin ladder system with metallic behaviour, which evidences the possibility of magnetically-mediated carrier pairing mechanism in a two-leg spin ladder lattice.

[*] Corresponding author, email: guochu.deng.cn@gmail.com

**I. Introduction**

After several decades of theoretical and experimental research, our understanding of the one-dimensional (1D) magnetic system has been significantly extended, partially thanks to the improvements in numerical and experimental methods.[1] However, similar progress has not occurred in the field of two-dimensional (2D) magnetic models. This is true particularly for the pairing mechanism of high temperature superconductors (HTSCs), which is widely believed to arise from the properties of holes in 2D spin lattices. Nearly three decades after the discovery of HTSC, the understanding to HTSC is still incomplete even though abundant intriguing experimental results have been collected. The challenge arises from various aspects, such as the complexities of the problem itself, but also from the limits of existing many-body techniques (including numerical methods in 2D system)[1]. Spin ladders are a finite number of coupled spin chains. They are essentially quasi-one-dimensional spin systems and can be extended into 2D with increasing numbers of spin legs. They are ideal model materials to study the transition from 1D to 2D spin lattice, since they have similarities to 2D spin lattices in HSTC.[1-4]

The 1D-2D bridging effect of spin ladders attracted extensive research concerns in the field of condensed matter physics, resulting in a rapid progress in the understanding of various types of spin ladder systems. For example, the spin-gap/gapless prediction about spin-half even and odd-leg ladders made by Dagotto and Rice [5] with numerical method, was verified by the experimental work in $SrCu_2O_3$ and $Sr_2Cu_3O_5$ by Azuma et al.[6] At the same time, a hole pairing mechanism for spin ladder superconductivity was theoretically proposed by the same group[3]. Although its underlying pairing mechanism is still under debate, the subsequent discovery of superconductivity in the spin ladder compound $Sr_{14-x}Ca_xCu_{24}O_{41}$ (x>10) was an important development in this field and aroused intensive research interests [7].

$Sr_{14-x}Ca_xCu_{24}O_{41}$ (for a specific case, denoted as Ca$x$) is obtained through doping of Ca into its parent compound $Sr_{14}Cu_{24}O_{41}$ (denoted as Sr14). It consists of two magnetic Cu sublattices, namely, the spin chain sublattice and the two-leg spin ladder sublattice. Alkane earth metal cations are intercalated between these two sublattices. The two one/quasi-one-dimensional spin lattices run along *c* axis and are incommensurate with respect to each other. Interestingly, this system is an inherently self-hole-doped system, containing 6 holes in one formula unit [7]. The parent compound is easy to synthesize and has almost all 6 holes on chain sites. Its physics was widely investigated and is well understood. It has a charge density wave ground state with spin dimers on both chains and ladders. Two chain spins form an AFM dimer with a localized-hole-induced Zhang-Rice singlet in between while ladder spins dimerize along rungs, as in the pure two-leg spin ladder $SrCu_2O_3$. When gradually substituting Ca for Sr, the resistivity of $Sr_{14-x}Ca_xCu_{24}O_{41}$ decreases mainly due to the redistribution of holes from chains to ladders[7]. It is believed that the holes are localized on chains and become mobile when transferring to ladders. In the compounds with $x>10$, a superconducting phase was observed under hydrostatic pressure ranging from 3 to 8GPa [8]. The superconducting behaviour is widely believed to originate from the ladder sublattice. Although some effort has been made to study the dynamics in the superconducting phase under pressure, the results have been inconclusive. This is partly because, with high Ca doping, the physics of both the chain and ladder systems becomes much more complicated. Up to now, the ground state of this spin ladder superconductor in ambient pressure has not been well understood because of contradictive reports from different experimental methods.[9, 10] Similar to the high temperature superconducting cuprates with 2D lattices, investigating the physics of the normal phase of $Sr_{14-x}Ca_xCu_{24}O_{41}$ should provide a solid basis to understand the superconducting mechanism. Therefore, it is extremely important to reveal the physics of the normal state of this spin ladder compound in order to find out the differences between $Sr_{14-x}Ca_xCu_{24}O_{41}$ and other 2D superconductors, and further our understanding to their superconducting mechanisms.

In this study, we systematically investigated the spin dynamics of the whole series of $Sr_{14-x}Ca_xCu_{24}O_{41}$ by inelastic neutron scattering. It is discovered that the singlet ground state survives in all these doped compounds with *x* up to 13. We discuss the chemical disorder effect and the hole doping effect on the spin gap excitation when increasing *x*. Different hole behaviours are discussed in three different doping ranges depending on their effects. The reason for losing Q dependency of the excitation intensity in the highly doped compounds was also discussed. In the end, an updated phase diagram of this system was presented on the basis of the results from this work.

**II. Experimental**

Large single crystal samples of $Sr_{14-x}Ca_xCu_{24}O_{41}$ were grown using a commercial travelling solvent floating zone furnace, which was modified to sustain a high background pressure (up to 35bar) for the highly Ca-doped compounds. The crystal growth procedure was described in details in the reference[11]. The inelastic neutron scattering experiments were carried out on the triple axis spectrometer 1T1 at the Orphée reactor, Laboratoire Léon Brillouin (LLB) and TAIPAN at the Open Pool Australian Lightwater (OPAL) reactor, ANSTO.[12] On both instruments, PG (002) crystals were used for both the monochromators and analysers. The final energy was fixed at 14.7meV or 30.5meV (for some high energy transfer near 40meV) with a PG filter placed after samples to remove the high order contaminations. The open collimation was used for the inelastic scattering experiments on 1T1 with double focused monochromator and analyser. The single crystal samples, weighting from 1 to 2g, were aligned with the *a\** and *c\** reciprocal lattice vectors in the scattering plane. Most of the measurements were carried out near the antiferromagnetic (AFM) zone centre (4.5, 0, 3.5) or (1.5, 0, 10.5) in a configuration of an average unit cell with *a*~11.5Å, *b*~12.8 Å, *c*~27.5Å (~$7c_{ladder} \approx 10c_{chain}$), depending on the compositions. The inelastic intensities of different samples were normalized by the intensities of the Bragg peak (0014).

**III. Results**

For the convenience of comparison, we measured the parent compound Sr14 first. A strong spin gap excitation peak was observed at the AFM zone centre Q(4.5, 0, 3.5) (not shown here). By fitting to a Gaussian, the resonance-like peak centres at 32.5meV with a full width at half maximum (FWHM) 2.03meV. As we know from the previous studies,[13, 14] this peak arises from the ladder sublattice and reflects to the singlet-triplet spin gap excitation of the spin dimers. The spin gap observed in our work agrees very well with the previous reported values from the high energy chopper spectrometer by R. S. Eccleston et al.[13] and the triple axis spectrometer by J. E. Lorenzo et al.[14] Constant-E scans were also carried out along both $Q_H$ and $Q_L$ directions at energy transfer 32meV, confirming the reported features.

The observed $Q_H$ dependency from Sr14 is quite interesting and is plotted in Fig. 1b as the red circles. Two maximal intensities around $Q_H$=4.5 and $Q_H$=7 are visible in this curve. The sinusoidal intensity modulation with $Q_H$ suggests that it could be attributed to the dimer dynamic structure factor. According to the dynamical structure factor formulated by M. Müller et al.,[15] the integrated intensity of a dimerized spin chain or ladder has the following relationship with the wave vector *q* along certain direction given by:

$$I(\boldsymbol{q}) = A * \sin^2\left(\frac{qd}{2}\right) + B \qquad (1)$$

where *A* is a factor determined by the interdimer-intradimer exchange ratio and the separation interdimer along chain direction, *d* is the separation of the spin sites in one dimer, *B* is a *q* independent contribution. The measured $Q_H$ scan in Fig.2b can be well fitted to Equation (1). The fitted *d* value is ~3.82Å along *a* axis, which is highly consistent with the Cu-Cu distance along rungs in the parent compound. In addition, the periodicity from the above data is 3 r.l.u. along $Q_H$, corresponding to the Cu-Cu distance in the real space along *a* axis. These facts confirm our speculation that the intensity modulation originates from the structure factor of the ladders. As we will see in the latter section, Ca doping has a strong impact on $Q_H$ dependency of magnetic scattering intensity.

Very similar experiments were carried out in Ca3 and the data are plotted in Fig. 1a. The line shape clearly shows a sharp peak with a rather strong shoulder at the higher energy side. This looks very similar to that from the parent compound. A Gaussian fit to the peak in $x$=3 gives a peak at a higher energy transfer 33.6meV with an enhanced FWHM 2.3meV. This spin gap energy is close to the value (34.9meV) reported in the polycrystalline Ca2.8 sample by R. S. Eccleston et al.[13] The spin gap increase in both samples indicates that slight Ca doping can improve the spin ladder coupling exchange energy ratio $\alpha=J_{rung}/J_{leg}$ in this ladder system, which is proportional to the spin ladder gap. The $Q_H$ dependency in $x$=3 is plotted as the blue curve in Fig. 1b, exhibiting a very similar modulation feature as the parent compound.

From above, we know that the line shapes of the excitation spectrum in $x = 0$ and 3 are very similar. The data from $x$=7 shows substantial differences from the two samples above. Fig. 2a plots the energy scans at various temperatures in $x$=7. There is no strong peak showing up in the energy scan at 3.6K. With careful comparison to the high temperature scan at 220K, a very broad distribution of magnetic scattering signals can be detected in the energy range from 20meV to 38meV. The discrepancy between the scans at the peak centre ($Q_L$=3.5) and off-centre ($Q_L$=4 and 6) provide supports to the similar conclusion (see the inset in Fig.2b).

The most obvious feature of the scans at 3.6K and 220K is the peak at ~29meV, which is marked as P1 and P2 for 3.6 and 220K in the plot, respectively. Even though this peak survived at 220K, we could not simply exclude its magnetic origin because the magnetic coupling can survive up to room temperature due to both $J_{rung}$ and $J_{leg}$ >300K.[16] In order to clarify its origin, the subtracted intensity from the two scans at low and high temperatures was plotted in Fig. 2b. In this plot, there is no peak at all at the energy near 29meV, indicating that P2 is a nonmagnetic spurious peak. As pointed out by P. Bourges et al.,[17] the temperature dependency of any nuclear contribution is only expected to obey the standard temperature factor $[1-\exp(-\hbar\omega/k_BT)]^{-1}$. In the energy range of P2, the factor is very close to unity for temperature lower than 220K. Thus, P2 could be a phonon and its intensity didn't change much at two temperatures.

From the subtracted intensity in Fig.2b, we observed a maximal intensity at ~35.5meV while the magnetic scattering extends to 20meV. The subtracted magnetic intensity can be fitted to a Gaussian with the peak centre at 35.5meV and FWHM 17.9meV, as shown by the blue solid line in Fig.2b. Quite different from the scans in $x=0$ and 3, the excitation peak in $x=7$ is distributed in a very wide range from ~20meV to 35.5meV. As we know, these compounds have incommensurate crystal structures, and Ca and Sr are disordered on A sites. The local lattice distortion could be induced by Sr/Ca disorder, resulting in a wide distribution of Cu-Cu distances along both rungs and legs. Correspondingly, the spin gap $E_g$, proportional to the coupling exchange energy ratio $\alpha$, presents a very broad feature.

Fig.3a plots the energy scans at Q(4.5, 0, 3.5) and Q(4.5, 0, 6) in $x=11$. A sharp peak is observed at ~31meV at the zone centre. In Fig. 3b, the scans were conducted through another zone centre Q(4.5, 0, -3.5) at 3.6K and 220K, confirming the peak is magnetic. Both the peaks can be fitted with a Gaussian form, which gives the peak centre at 31.3meV with FWHM 4.65meV. Comparing to Ca7, the Ca11 peak is much sharper and its height recovers to some extent. Since the holes are gradually redistributed to the ladders on increasing $x$, we could not attribute the peak broadening in Ca7 to the increase of hole number because the Ca11 peak is not as broad as Ca7 and its height recovers. With half Ca/Sr on A sites, Ca7 has the strongest chemical disorder in the whole series, and shows the most significant broadening effect. It is natural to speculate that the disorder-induced structure distortion causes the wide distribution of the spin gap energy. Such an effect not only broadens but also slightly increases the excitation energy. Further increasing $x$ reduces chemical disorder and suppresses the structural distortion. Consequently, the peak becomes sharper again in $x=11$.

As for Ca12.2, the energy scans were performed at another zone centre Q(1.5, 0, 10.5) rather than Q(4.5, 0, 3.5) because of the higher intensity at the former. The spin gap peak of Ca12.2 is not obvious in Fig. 4a due to its reduced intensity and the complex background. By subtracting the high temperature background, we found the spin gap excitation peak, which is fitted with $E_g$=31.98meV and FWHM 4.67meV. It is clear that the Ca12.2 peak is much weaker than the Ca11 one. However, their FWHMs are very similar.

The most interesting observation in $x$=12.2 is no $Q_H$ dependency of the magnetic scattering intensity, completely different from the results in $x$=0 and 3. As shown in Fig. 4b, the counts is higher in the whole range (0.9<$Q_H$<2.5) at $Q_L$=10.5 than at at $Q_L$=9.75. The former is through AFM zone centre while the latter is away from it. The intensity subtraction does not show any peak at all. In the whole doping series, it is clear that intensity modulation along $Q_H$ gradually evolves from strong ($x$=0, 3) to weak ($x$=7, 11) and finally develops into $Q_H$ independency ($x$=12.2). This tendency hints that the magnetism of the spin ladder lattice undergoes a substantial change when increasing $x$. We will discuss this in more details later.

Very similar to Ca12.2, Ca13 has a weak excitation peak. $Q_L$ scans performed at a series of energy transfers were plotted in Fig. 5a and b. In this plot, no obvious peak is resolved at 28meV while a peak shows up at 30meV. The maximum intensity from the spin gap excitation was observed at 32meV and then it gradually decreased with the increase of energy.

## IV. Discussion

All the experimental results are summarized in Fig. 6. It plots the spin gap energy $E_g$, the FWHMs and the integrated intensities against the Ca doping content in $Sr_{14-x}Ca_xCu_{24}O_{41}$. The first important information from Fig. 6 is that the spin gap excitation was observed in all these compounds, indicating a singlet ground state nature of the whole series. This refutes the conclusion that the spin gap gradually closes and collapses at a certain Ca doping level by using NMR.[9] In addition, we also find that the spin gap slightly increases with $x$ when $x$<7, while it is nearly constant for $x$≥11. It is interesting to note that the spin gap for $x$=0 and 13 are nearly the same. The half FWHMs, which are marked as the hatched blue bars below the excitation energy, are around 1 ~ 2 meV at both the low and high doping ends. In contrast, it is much larger (~8meV) in the half doping compound Ca7. We attribute this broadening effect to chemical disorder.

Concerning the integrated intensity, it slightly increases with $x$ in the range of $x<7$, then decreases from $x=7$ to 11 and suddenly drops at $x=12.2$. We notice that the integrated intensity in $x=7$ is even higher than those in $x=0$ and 3 even though the peak of Ca7 is severely damped. Now the question is how to understand the $x$ dependency of the integrated intensity. As we know, it has been established that the doped holes gradually redistribute from the chain sublattice to the ladder sublattice on increasing $x$.[18, 19] Apparently, the continuous increase of hole number on the ladder sublattice from $x=0$ to 13 is not consistent with the first-increase and then-decrease tendency of the integrated intensity in the whole series. If the hole number on the ladder is not able to account for the intensity drop, we need to shift our attention to which roles those holes play in these magnets. M. Kato et al. reported the electric transport properties in $Sr_{14-x}Ca_xCu_{24}O_{41}$ with $0 \leq x \leq 9$.[20] They observed that the temperature dependence of the thermoelectric power changed from negative to positive when $x$ increases from 6 to 8.4, which was considered by the authors as an evidence of the insulating metallic transition.[20] The $x$ range for such a transition is consistent with the turning point of the intensity dependency on $x$ in Fig. 6. On the basis of this fact, we speculate that it is not the hole number but the hole delocalization on the ladders that causes the suppression of the intensity. Another resistivity measurement from N. Motoyama et al. indicated that the temperature dependency of resistivity along $c$ axis becomes positive when $x \geq 11$, indicating a metallic property.[21] Additionally, we had found that the Cu valence modulation in the ladder sublattice becomes less pronounced in the range of $x>11$, meaning the hole delocalization in these compounds.[18]. Therefore, the sudden drop of the intensity in $x=12.2$ and 13 must be due to the metallic behaviour of the ladders.

With the analysis above, we can depict a picture of hole behaviour in the whole doping range. For $x\leq 7$, holes are still localized on the ladder at low temperature. They have very limited impact on the singlet dimer ground state of the ladder. In the range of $7<x\leq 11$, holes partially become delocalized on the ladder, breaking singlet dimmers to some extent and suppressing their excitation intensity. Further increasing $x$ above 11, the ladder sublattice becomes metallic, and the magnetic behaviour has been fundamentally changed. Such a change causes the extensive depression of the peak intensity at $x$=12.2 and 13.

Another interesting observation is the loss of $Q_H$ dependency of the excitation in $x$=12.2. As discussed above, the intensity modulation along $Q_H$ has been clearly observed in $x$=0, 3 and 7 (even $x$=11, but weaker), which can be attributed to the dimer structure factor. However, the $Q_H$-independent scattering in Ca12.2 indicates a nonlocalized spin coupling in the rung direction. The changes of intensity modulation could be explained by a scenario of the crossover from a local Cu moment magnetism to an itinerant hole magnetism on increasing $x$. That is, the spin exchange is between localized $Cu^{2+}$ moments in $x\leq11$ while it takes place between the itinerant holes in $x>11$. As known in metallic YBCO, the spin correlation length is shorter than a few lattice distances, much less than that in the insulating phase, resulting in the decoupling of the spins in different bilayers.[22] Similarly, when the spin ladders transform from an insulating state into an itinerant metallic state, both the Cu-Cu spin correlation length along the leg and rung directions are substantially reduced. Since the coupling along $Q_H$ is much weaker than that along $Q_L$, the hole doping effect has much stronger impact on the correlation along $Q_H$. Due to the decoupling of the neighbour spin ladders, the dynamic structure factor is an incoherent superposition of the signals arising from individual spin ladders, resulting a randomly distributed scattering intensity along $Q_H$. Thus, we could understand the strong intensity modulation in Sr14, but a $Q_H$-independent feature in Ca12.2. The excitation intensity substantially drops in $x$=12.2 because the magnetic scattering signals are averagely distributed along $Q_H$.

As well known, a gapped excitation (also called magnetic resonance peak) in the heavily-doped metallic YBCOs were only observed below $T_c$, and is believed related to the pairing mechanism in the superconducting state. Surprisingly, we observed the spin gap excitation in the normal metallic state of this spin ladder series with $x>11$. Even though it is too early to discuss the relationship between this excitation and the superconducting phase under pressure in these compounds, our observation hints the possibility of hole pairing mechanism mediated by magnetic interaction in an itinerant magnetic system. Further experiments are definitely needed to search for the resonance excitation in the superconducting state of this spin ladder compound and compare the origins of two excitations in the normal and superconducting states.

**V. Conclusion**

In conclusion, the spin gap of $Sr_{14-x}Ca_xCu_{24}O_{41}$ was observed in all the compositions up to $x$=13 by using inelastic neutron scattering. The gap energies slightly change while the peak intensities strongly evolve with $x$. We found that chemical disorder enormously broadens the excitation peak, and the delocalization of holes on the ladder suppresses the peak intensity. Losing $Q_H$ dependency in $x$=12.2 is attributed to the development of an itinerant magnetism. The survival of the spin gap in $x$=12.2 and 13 suggests the possibility for magnetic exchange to mediate hole pairing behaviour in an itinerant magnetic system. In order to clarify the relationship between this spin gap and the superconducting phase, further direct measurement of the superconducting resonance excitation is necessary and will be continued in the future.

**Acknowledgement**

The experiment at LLB was supported by the European Commission under the 6-th Framework Programme through the Key Action: Strengthening the European Research Area, Research Infrastructures (Contract No.: RII3-CT-2003-505925 (NMI3)). Some of the authors gratefully acknowledge the financial support from the Indo-Swiss Joint Research Program (ISJRP, Contract No. JRP122960) by Swiss State Secretariat of Education and Research.

**Figures and Legends**

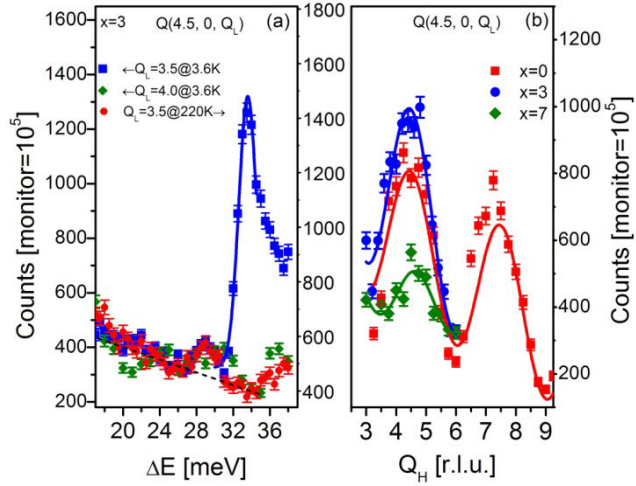

Fig. 1 (a) Energy scans at different temperature and different $Q_L$ from $Sr_{11}Ca_3Cu_{24}O_{41}$; The sharp resonace-like peak is fitted to a Gaussian. The green curve is an off-center background scan and the read curve is a scan at 220K; The black dash line is given as guide to the eye. (b) $Q_H$ dependency of the peak intensities in $Sr_{14-x}Ca_xCu_{24}O_{41}$ with $x$=0, 3, 7. The solid lines are fitted to the experimental data by using Eq (1) (see text).

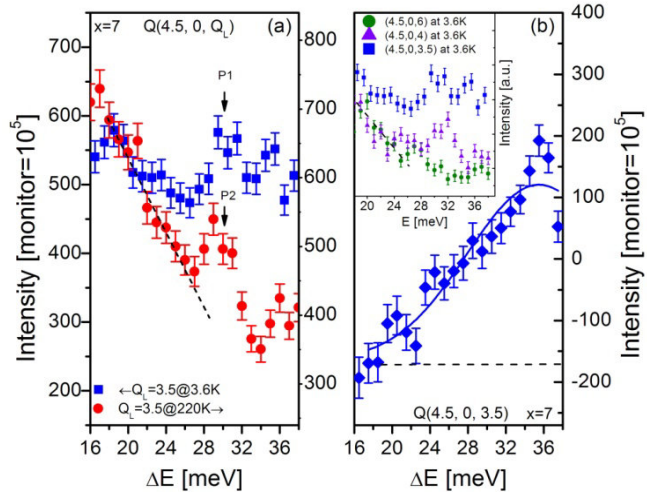

Fig. 2 (a) Energy scans at 3.6K and 220K from $Sr_7Ca_7Cu_{24}O_{41}$. P1 and P2 correspond to the same spurious peak at 3.6K and 220K, respectively. The arrows (← and →) in the legend pointed out the left or right scale corresponding to the curves. The dash line is a guide to the eye as the background. (b) The subtracted intensity between 3.6K and 220K, the solid curve is fitted to a Gaussian and the dash line is a guide to the eye as the background. The inset is the energy scans at $Q_L$=3.5, 4 and 6 from $Sr_7Ca_7Cu_{24}O_{41}$. These results indicate that the magnetic scattering extends to ~20meV.

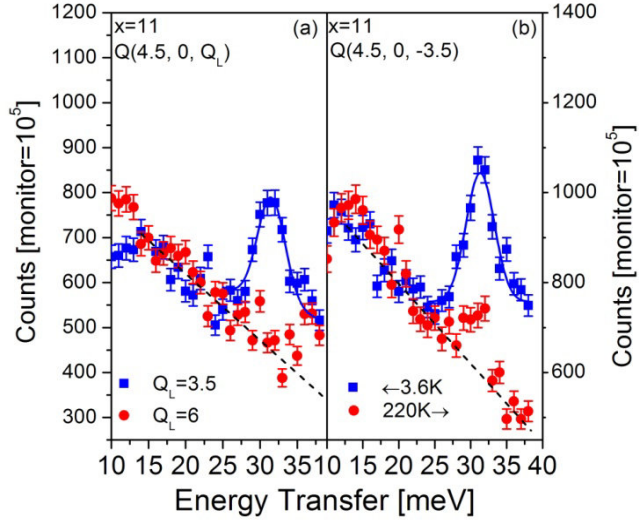

Fig. 3 (a) Energy scans at Q(4.5, 0, 3.5) and Q(4.5, 0, 6) from $Sr_3Ca_{11}Cu_{24}O_{41}$ at 3.6K; (b) Energy scans at Q(4.5, 0, -3.5) at 3.6K and 220K from $Sr_3Ca_{11}Cu_{24}O_{41}$. The left and right arrows point the scales used for the corresponding curves. The solid lines are fitted to Gaussians and the dash lines in both (a) and (b) are guides to the eye.

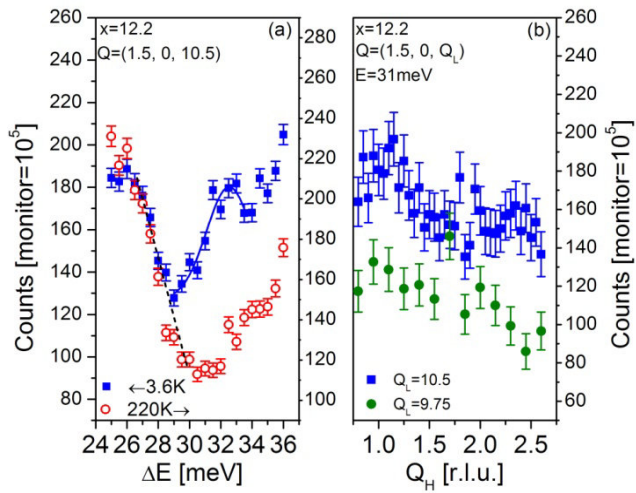

Fig. 4 (a) Energy scans at Q(1.5, 0, $Q_L$) at 3.6 and 220K from $Sr_{1.8}Ca_{12.2}Cu_{24}O_{41}$; (b) $Q_H$ scans at Q(1.5, 0, $Q_L$) at $Q_L$=10.5 and 9.75, corresponding to the centre and off-centre peak positions, respectively. The left and right arrows indicate the scales for each curve. The solid line in (a) is a Gaussian fitting and the dash line in (a) is a guide to the eye.

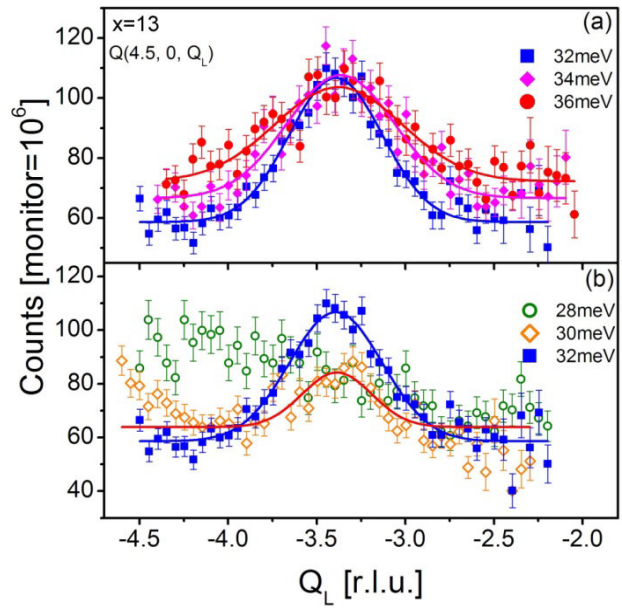

Fig. 5 $Q_L$ scan near Q=(4.5, 0, $Q_L$) at 36, 34, 32meV (a) and 32, 30 and 28meV (b) in $SrCa_{13}Cu_{24}O_{41}$ at 2K. The solid lines are fitted to Gaussians.

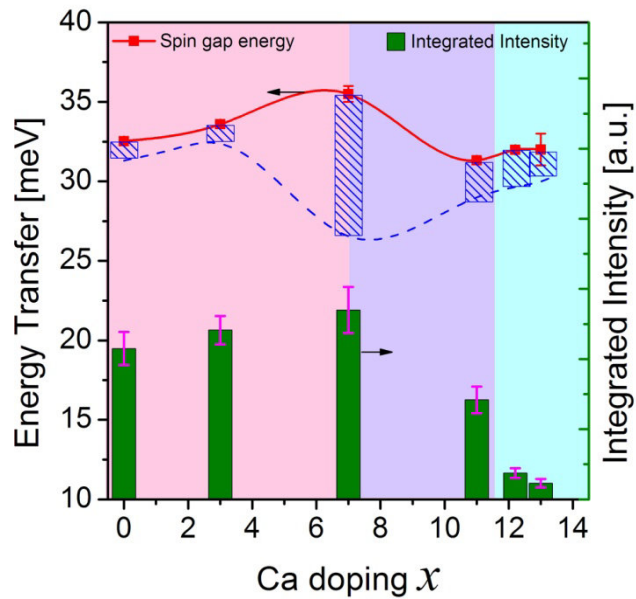

Fig. 6 $x$ dependencies of the spin gap excitation energy (read square), integrated peak intensity (green bars) with error bars in $Sr_{14-x}Ca_xCu_{24}O_{41}$. The blue hatched bars show the half FWHM of the fitted resonance-like peaks at each $x$, indicating the broadening effect of doping.

The red solid and blue dash lines are just guides for the eye. The pink area denotes the insulating range, where local Cu moment magnetism is predominant. The purple denotes an insulating–metallic crossover region. And the cyan area denotes the metallic state, where the itinerant magnetism is overwhelming.